\documentclass[sigconf]{acmart}

\AtBeginDocument{%
  }

\usepackage{amsmath,amssymb,amsfonts}
\usepackage{algorithmic}
\usepackage{graphicx}
\usepackage{textcomp}
\usepackage{xcolor}
\usepackage{url} 
\usepackage{multirow}
\usepackage{booktabs}
\usepackage{tcolorbox}
\usepackage{diagbox}
\usepackage{tabularx}
\usepackage{microtype}
\usepackage{stackengine}  
\usepackage{listings}
\usepackage{subcaption} 
\usepackage{stackengine}
\usepackage{amsmath}
\usepackage{algorithm}
\usepackage{algorithmic}
\usepackage{wrapfig}
\usepackage[table]{xcolor} 
\usepackage{colortbl}

\newcommand{\findingboxx}[1]{
\begin{center}
\begin{tcolorbox}[colback=gray!11,
                  colframe=black,
                  boxrule=0.2mm,
                  width=0.45\textwidth,
                  arc=.8mm, auto outer arc,
                 ]
 #1
\end{tcolorbox}
\end{center}}

\setcopyright{acmlicensed}
\copyrightyear{2025}
\acmYear{2025}
\acmDOI{XXXXXXX.XXXXXXX}

\begin{document}

\title{Mathematically-Guided Detection of Floating-Point Errors}

\author{Youshuai Tan}
\affiliation{%
  \institution{The Hong Kong University of Science and Technology (Guangzhou)}
    \country{China}
}
\email{ytan387@connect.hkust-gz.edu.cn}

\author{Zhanwei Zhang}
\affiliation{%
  \institution{The Hong Kong University of Science and Technology (Guangzhou)}
    \country{China}
}
\email{itbill8888@gmail.com}

\author{Zishuo Ding}
\authornote{Corresponding author.}
\affiliation{%
  \institution{The Hong Kong University of Science and Technology (Guangzhou)}
  \country{China}
  }
\email{zishuoding@hkust-gz.edu.cn}

\author{Lianyu Zheng}
\affiliation{%
  \institution{The Hong Kong University of Science and Technology (Guangzhou)}
    \country{China}
}
\email{lianyuzheng81@gmail.com}

\author{Jinfu Chen}
\affiliation{%
  \institution{Wuhan University}
  \country{China}}
  \email{jinfuchen@whu.edu.cn}

\author{Weiyi Shang}
\affiliation{%
 \institution{University of Waterloo}
 \country{Canada}}
 \email{wshang@uwaterloo.ca}

\renewcommand{\shortauthors}{Trovato et al.}

\begin{abstract}
Floating-point program errors can lead to severe consequences, particularly in critical domains such as military applications. Only a small subset of inputs may induce substantial floating-point errors, prompting researchers to develop methods for identifying these error-inducing inputs. Although existing approaches have achieved some success, they still suffer from two major limitations: \textbf{(1) High computational cost:} The evaluation of error magnitude for candidate inputs relies on high-precision programs, which are prohibitively time-consuming. \textbf{(2) Limited long-range convergence capability:} Current methods exhibit inefficiency in search, making the process akin to finding a needle in a haystack. Even when the search probe approaches the vicinity of error-inducing inputs, existing methods can still fail to achieve convergence.

To address these two limitations, we propose a novel method, named MGDE, to detect error-inducing inputs based on mathematical guidance. With respect to the first limitation, we avoid using errors computed by high-precision programs as indicators during the search process. Instead, we identify inputs that lead to a large condition number of a specific operation. For the second limitation---a fundamental challenge common to all existing methods---we reformulate scenarios with large condition numbers into an equality constraint. This transforms the task of detecting error-inducing inputs into a root-finding problem. By employing the Newton-Raphson method, which exhibits quadratic convergence properties, we achieve highly effective and efficient results. Since the goal of identifying error-inducing inputs is to uncover the underlying bugs, we use the number of bugs detected in floating-point programs as the primary evaluation metric in our experiments. As FPCC represents the most effective state-of-the-art approach to date, we use it as the baseline for comparison. The dataset of FPCC consists of 88 single-input floating-point programs. FPCC is able to detect 48 bugs across 29 programs, whereas our method successfully identifies 89 bugs across 44 programs. Moreover, FPCC takes 6.4096 times as long as our proposed method. We also deploy our method to multi-input programs, identifying a total of nine bugs with an average detection time of 0.6443 seconds per program. In contrast, FPCC fails to detect any bugs while requiring an average computation time of 100 seconds per program.

\end{abstract}




\setcopyright{none} 
\settopmatter{printacmref=false} 
\maketitle

\section{Introduction}
\label{Introduction}
Floating-point computations serve as the foundation for contemporary scientific and engineering applications, where inaccuracies can have severe real-world impacts. These errors have been documented in critical domains such as military operations~\cite{skeel1992roundoff}, air and space engineering~\cite{lions1996flight}, and economic infrastructures~\cite{weisstein1999roundoff}. Therefore, extensive research has been devoted to detecting errors in floating-point programs~\cite{nethercote2007valgrind, benz2012dynamic, franccois2016verrou, sanchez2018finding, panchekha2015automatically, bao2013fly, lee2015raive, chowdhary2021parallel, chowdhary2022fast}. One popular research direction focuses on identifying error-inducing inputs~\cite{chiang2014efficient, zou2015genetic, yi2019efficient, wang2022detecting, guo2020efficient, zhang2023eiffel, zhang2024hierarchical, zou2019detecting, yi2024fpcc}, as only a small subset of inputs tends to cause substantial errors in floating-point programs~\cite{bao2013fly}. Therefore, detecting such inputs is essential for ensuring reliability for floating-point programs.

Although existing approaches for detecting error-inducing inputs have made significant progress, they still suffer from two major limitations. \textbf{Limitation 1: Computing errors with high-precision programs is time-consuming.} A prerequisite for identifying error-inducing inputs is the ability to compute the error corresponding to different inputs. Most existing tools treat the output of a high-precision program as the oracle for error computation. However, such high-precision programs often require significantly more time to execute---in some cases over 100 times longer---thereby severely limiting the efficiency of input-space exploration. \textbf{Limitation 2: Existing tools are unable to support long-range convergence.} All existing approaches can essentially be regarded as search algorithms. However, they are inherently heuristic in nature and are only capable of identifying error-inducing inputs when their probes lie close to the correct solutions. They fail to support long-range convergence, i.e., discovering error-inducing inputs starting from regions that are far from the ground truth. Consequently, the search procedure requires an extremely dense placement of probes. However, given the vastness of the double-precision space, these approaches often resemble searching for a needle in a haystack. Although some tools employ optimization techniques such as evolutionary algorithms~\cite{zou2019detecting} or the direct search algorithm~\cite{gablonsky2001locally}, they are still fundamentally heuristic and do not escape this limitation. 

To overcome these limitations, we propose a novel method, named MGDE (Mathematically-Guided Detection of floating-point Errors), to detect error-inducing inputs. To avoid relying on errors computed by high-precision programs during the search process, ATOMU~\cite{zou2019detecting} proposes transforming the problem of searching for error-inducing inputs into the task of identifying inputs that maximize the condition number of a specific atomic operation (e.g., addition, subtraction, \(\sin\), \(\tan\)) in the floating-point program. According to condition number theory, a large condition number for an operation typically leads to significant computational errors. Our approach adopts this strategy to eliminate the need for high-precision computation, thereby reducing the runtime overhead of our method. Although ATOMU avoids relying on high-precision programs, its search strategy remains heuristic in nature. As a result, it can only succeed when the probe is very close to the ground truth, indicating a lack of long-range convergence. Ideally, we envision a method that, akin to gradient descent in machine learning, can effectively guide the search toward the desired error-inducing inputs. Therefore, we transform the problem of searching for inputs that cause the condition number of a certain operation to increase into a root-finding problem. This is because, when the condition number becomes sufficiently large, the result of the operation may converge to a specific value or tend toward infinity. In the first scenario, we can directly set the result of the operation equal to that specific value, thereby constructing an equation. In the second scenario, we can set the reciprocal of the operation's result to zero, which also allows us to formulate an equation. This approach enables us to employ the Newton-Raphson method for rapid and long-range convergence search.

Since the objective of identifying error-inducing inputs is to detect bugs, we adopted the number of triggered bugs as our evaluation metric. Experimental results demonstrate that for 88 single-input programs, our method detects 89 bugs across 44 programs, while FPCC identifies only 48 bugs in 29 programs. Our method exhibits a 31.7941× higher bug discovery rate per second compared to FPCC. For multi-input programs, our method successfully detects nine bugs, whereas FPCC fails to identify any. Furthermore, our average processing time per program is merely 0.6443 seconds, compared to FPCC's 100 seconds. These results demonstrate that our method achieves remarkable performance due to its long-range convergence capability, highlighting its strong practical value. The replication package is available on the following link\footnote{https://github.com/anonymous20250719/MGDE}.

In summary, our work makes the three contributions:
\begin{itemize}
\item We propose MGDE, a novel method designed to efficiently and effectively detect error-inducing inputs.
\item We evaluate our method on both single-input and multi-input floating-point programs.
\item Our method uniquely achieves long-range convergence, inspiring future studies to leverage rigorous mathematical foundations for enhanced detection.
\end{itemize}

\textbf{Paper organization.} The rest of the work is structured as follows. In Section~\ref{bac}, we provide the necessary background for our study. Section~\ref{motivate} discusses the shortcomings of existing methods. We detail our proposed method in Section~\ref{method}. Section~\ref{eva} presents the evaluation results. We discuss potential limitations in Section~\ref{threat} and review related works in Section~\ref{related-work}. Finally, Section~\ref{conclusion} summarizes our contributions and concludes the paper.

\section{Background}
\label{bac}
In this section, we introduce the condition number theory and the Newton–Raphson Method. Moreover, we explain how we use them in our method.

\subsection{Condition number}
\label{condition_number}
The condition number is an essential metric that quantifies the amplification of input variations through the function \(f(x)\)~\cite{overton2001numerical}. The formula can be mathematically derived by applying the Taylor Expansion Theorem~\cite{zou2019detecting}. The relative error of \(f(x)\) with error \(\Delta x\) can be derived as:

\begin{equation}\begin{aligned}
&=\left|\frac{f(x+\Delta x)-f(x)}{f(x)}\right| \\
&=\left|\frac{f(x+\Delta x)-f(x)}{\Delta x}\cdot\frac{\Delta x}{f(x)}\right| \\
&=\left|(f^{\prime}(x)+\frac{f^{\prime\prime}(x+\theta\Delta x)}{2!}\Delta x)\cdot\frac{\Delta x}{f(x)}\right|,\theta\in(0,1) \\
&=\left|\frac{\Delta x}{x}\right|\cdot\left|\frac{xf^{\prime}(x)}{f(x)}\right|+O\big((\Delta x)^{2}\big) \\
&=Err_{rel}(x,x+\Delta x)\cdot\left|\frac{xf^{\prime}(x)}{f(x)}\right|+O\big((\Delta x)^2\big)
\end{aligned}
\end{equation}
where \(\theta\) is the Lagrange form of the remainder. As the term \(O\big((\Delta x)^2\big)\) can be discarded, we obtain the mathematical expression of the condition number: \(\left|\frac{xf^{\prime}(x)}{f(x)}\right|\). The mathematical implication of the condition number is that if a function has a large condition number and the input contains errors, the resulting output will exhibit significant error amplification.

In the field of detecting error-inducing inputs, ATOMU~\cite{zou2019detecting} and FPCC~\cite{yi2024fpcc} employ the condition number as a fitness metric to search for such inputs. The theoretical basis for their approaches lies in the observation that significant errors in floating-point computations primarily originate from operations with large condition numbers, since the corresponding operands inevitably contain rounding errors. For example, concerning the expression \(\sin(x) - 0.4\), when x is 0.411516846067, the result of \(\sin(x)\) approaches 0.4. In the case of subtraction involving two nearly equal numbers, the derivative is 1. Thus, according to the formula for condition number, the condition number with respect to \(\sin(x)\) equals the operand divided by the result of this subtraction (a very small value), leading to an extremely large final condition number (\(\left|\frac{\sin(x)}{\sin(x)-0.4}\right|\)). Since the computed result of \(\sin(x)\) inevitably contains rounding errors---specifically, the double-precision evaluation yields \textbf{0.3999999999995527}, whereas higher-precision computation gives \textbf{0.3999999999995527}228457665138406588994966...---the error, though extremely small in \(\sin(x)\) itself, is amplified by the condition number. This leads to a relative error of \(7.09 \times 10^{-5}\) in the final result.

To enhance the efficiency and effectiveness of detection by employing non-heuristic algorithms, we reformulate the problem of identifying error-inducing inputs as a root-finding problem. For instance, in the aforementioned example, since we know that significant errors arise when \(\sin(x) - 0.4\) approaches zero, we directly set the equation \(\sin(x) - 0.4=0\). The solutions to this equation then yield the desired inputs.

\subsection{Newton-Raphson method}
\label{Newton-Raphson_method}

Around the 1670s, Newton developed a method for approximating roots of equations, which he illustrated using the example of the cubic equation \(x^3-2x-5=0\)~\cite{gowers2010princeton}. This method exhibited the following formal structure: \textbf{1) Initial condition specification:} Establishing an approximate solution \(x_0=2\) through empirical estimation.\textbf{ 2) First-order transformation:} Implementing the substitution \(x=x_0+p\), which generated the new equation: \(p^3+6p^2+10p-1=0\). \textbf{3) Linear approximation:} Since \(x\) is close to 2, the value of \(p\) is small. Therefore, he simplified the expression by neglecting the \(p^3\) and \(6p^2\) terms, as they are much smaller compared to the dominant term \(10p-1\). This simplification leads to the equation \(10p-1=0\), giving the approximate solution \(p=0.1\). While the exact solution is not generated, it offers a more accurate estimate, \(x_1\), which is now taken as 2.1. \textbf{4) Recursive refinement:} He then reinitialized the process with \(x=x_1+q\) and obtained \(q=-0.0054, x_2=2.0946\). Subsequently, he iteratively repeated this process until a result with sufficiently high accuracy is obtained.

\begin{figure}[htbp]
\vspace{-0.4cm}
    \centering
    \includegraphics[width=0.49\textwidth]{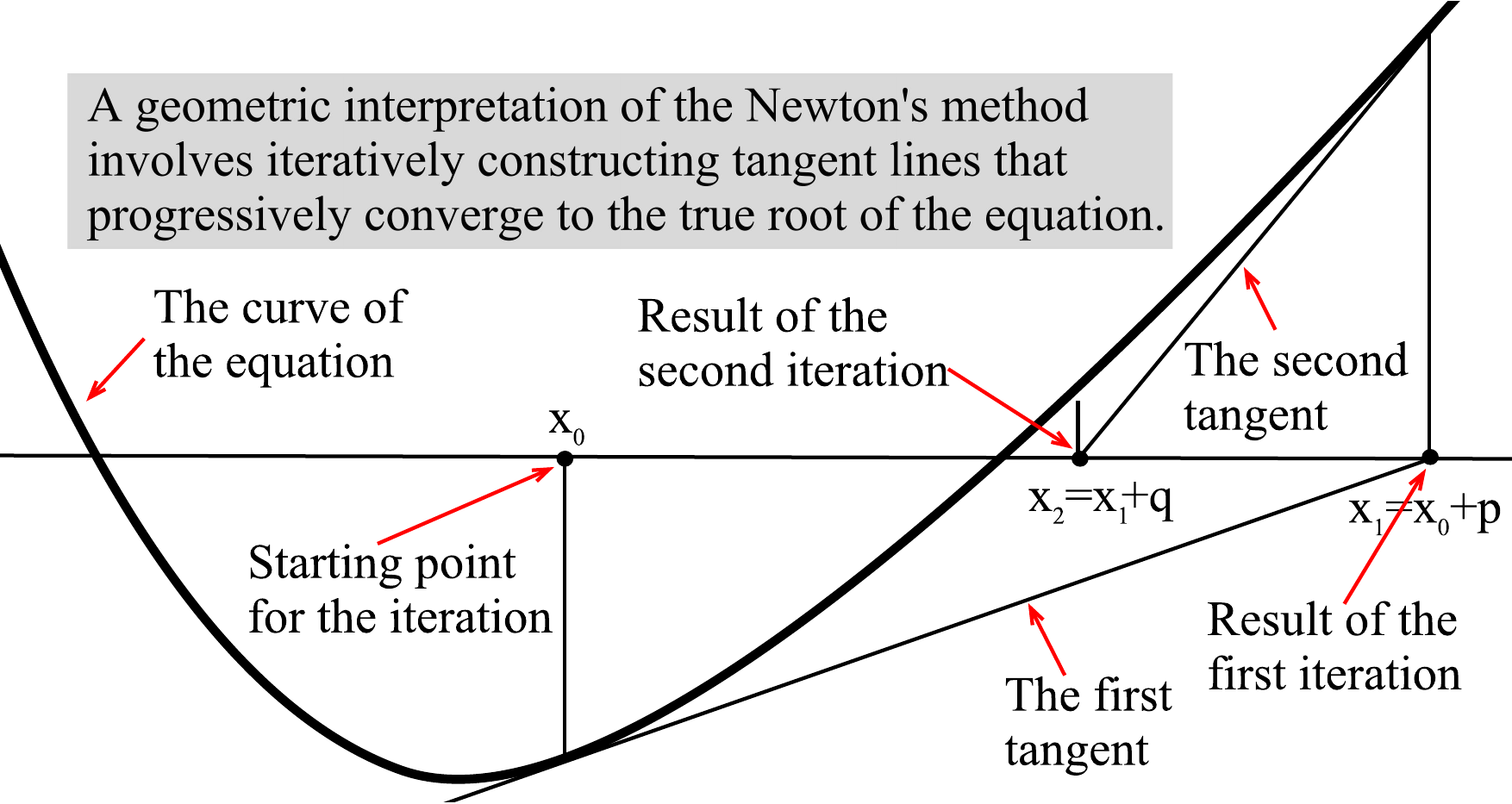}
    \vspace{-0.7cm}
    \caption{Geometric illustration of the Newton's method.}
    \vspace{-0.2cm}
    \label{newton}
\end{figure}

Newton's method can be interpreted geometrically through the graph of the equation \(f(x)=0\). A solution to the equation is where the curve intersects the x-axis. Suppose we begin with an initial guess \(x_0=2\) for the root, as Newton did above. We then substitute \(x=2+p\) into the original function \(f(x)\), transforming it into a function expressed in terms of \(p\): \(g(p)=p^3+6p^2+10p-1\). Since we discard higher-order terms and keep only the constant and linear ones, we easily obtain the value of \(x_1\). Geometrically speaking (cf. Figure~\ref{newton}), the function obtained by neglecting the higher-order terms represents the tangent line at \(p=0\), and the value \(x_1\) corresponds to the point where this tangent intersects the x-axis. Therefore, Newton's method is also known as the tangent method. As illustrated in Figure~\ref{newton}, whenever the new approximation point lies between the previous estimate and the true solution, it invariably yields a more accurate approximation. Moreover, the method exhibits \textit{quadratic convergence}, with each iteration's error being approximately proportional to the square of the previous one.

Although the aforementioned method can efficiently locate the root, each iteration requires constructing a new equation. This limitation was resolved by Raphson in 1690. He devised a universal formula applicable to every iteration step, which furthermore admits an elegant interpretation in terms of tangents. In Figure~\ref{newton}, the value of the first tangent can be calculated by: \((0-f(x_0))/(x_1-x_0)\). After transformation, we obtain: \(x_1 = x_0 - f(x_0)/f'(x_0)\). This equation enables us to iteratively compute subsequent approximations using a unified formula. The Newton-Raphson method derives directly from this fundamental formula.

We formulate the problem of detecting error-inducing inputs as a root-finding problem and employ the Newton-Raphson method to obtain solutions efficiently and effectively.
\section{Motivation}
\label{motivate}

In recent years, the search for error-inducing inputs has emerged as a highly active research topic in the fields of software engineering and programming languages~\cite{chiang2014efficient, zou2015genetic, yi2019efficient, wang2022detecting, guo2020efficient, zhang2023eiffel, zhang2024hierarchical, zou2019detecting, yi2024fpcc}. However, the effectiveness of these methods leaves room for improvement. For instance, the state-of-the-art algorithm, FPCC~\cite{yi2024fpcc}, successfully identifies error-inducing inputs in only 29 out of the 88 target functions. The suboptimal search performance can be explained by two limitations in the existing methods:

\textbf{Limitation 1: Using high-precision programs for error computation incurs excessive runtime overhead.} Previous approaches typically convert target programs into high-precision versions and treat them as oracles to compute errors. These errors are then used as indicators to guide subsequent searches. However, since high-precision computations incur significant runtime overhead, they severely degrade the efficiency of the search process, rendering it impractical for large-scale or performance-sensitive applications. Performance benchmarks show that quadruple-precision (128-bit) floating-point operations typically run approximately 100 times slower than their double-precision (64-bit) counterparts~\cite{larsson2013exploring}. The performance impact becomes even more pronounced when using arbitrary-precision libraries like MPFR~\cite{fousse2007mpfr} and \textit{mpmath}~\cite{mpmath}, where computational overhead grows significantly as precision requirements increase.

\textbf{Limitation 2: Fails to support long-range convergence.} Previous approaches have predominantly relied on heuristic methods, such as genetic algorithms, to search for error-inducing inputs. Although genetic algorithms demonstrate efficacy in certain scenarios, particularly where the derivative of the function is unavailable, they are ill-suited for this specific task. This is because error-inducing inputs only occur within an exceedingly narrow range, and the search space (double-precision floating-point numbers) is prohibitively vast. Without an efficient search strategy, this task becomes akin to searching for a needle in a haystack. In other words, although the search trajectories of these methods come very close to the error-inducing inputs, they still fail to converge, ultimately missing the correct result.

\begin{figure*}[htbp]
\vspace{-0.4cm}
    \centering
    \includegraphics[width=1\textwidth]{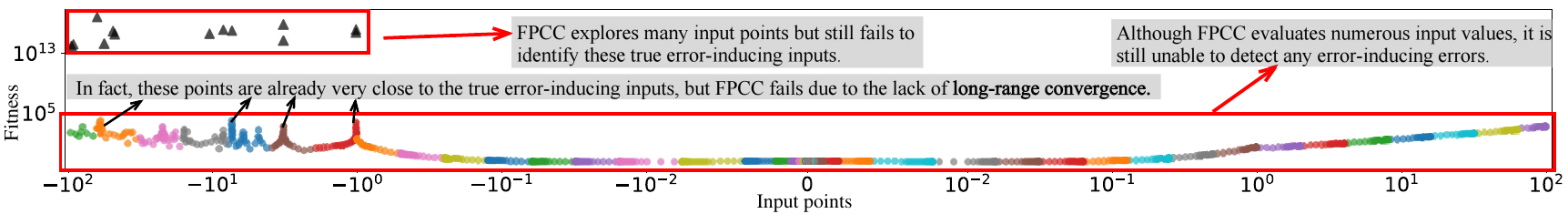}
    \vspace{-0.7cm}
    \caption{Detection process of FPCC on \textit{gsl\_sf\_airy\_Ai\_deriv\_e}. Black triangles denote the fitness values of the true error-inducing inputs missed by FPCC. Colored circles correspond to the fitness values of inputs evaluated by FPCC, with each color indicating a distinct region defined by FPCC’s partitioning scheme.}
    \vspace{-0.3cm}
    \label{trace_all_intervals}
\end{figure*}

To demonstrate the ineffectiveness of heuristic algorithms in addressing the problem of error-inducing inputs, we present a concrete case study. FPCC~\cite{yi2024fpcc} is currently the state-of-the-art method in this topic. However, despite replacing traditional high-precision oracle-computed errors with chain conditions to improve efficiency, it still fails to support long-range convergence. The FPCC dataset includes 88 single-input GNU Scientific Library (GSL)~\footnote{~\url{https://www.gnu.org/software/gsl/}} functions, among which error-inducing inputs were identified for 29 functions, excluding function \textit{gsl\_sf\_airy\_Ai\_deriv\_e}. Figure~\ref{trace_all_intervals} demonstrates the search trajectory of FPCC across the \([-100, 100]\) interval of function \textit{gsl\_sf\_airy\_Ai\_deriv\_e}. While approximately 2,860 sampling floating-point numbers are evaluated within this interval, FPCC fails to detect any valid error-inducing inputs. It should be noted that even after increasing FPCC's runtime by approximately 100-fold, no valid error-inducing inputs are identified. We attribute the difficulty of FPCC in detecting error-inducing inputs to its heuristic strategy, which exhibits insufficient long-range convergence.

To overcome these limitations, we introduce MGDE, a mathematically-guided method for detecting error-inducing inputs.

\section{MGDE: Detecting error-inducing inputs with the Newton-Raphson method}
\label{method}

Identifying error-inducing inputs of floating-point programs is an important and popular task. However, existing approaches rely on heuristic strategies and lack guarantees of long-range convergence. As a result, even when the probes are already close to the target error-inducing inputs, they may still fail to identify them (\textbf{limitation 2}). Another limitation of existing tools is their reliance on high-precision programs to compute errors during the detection process. The high computational cost of such programs significantly reduces the overall efficiency of the tools (\textbf{limitation 1}). To address the above limitations, we propose a novel method, named MGDE. First, we reframe the error-inducing inputs detection task as a root-finding problem. Second, we apply the Newton-Raphson method to this problem, taking advantage of its high convergence speed. Finally, since some candidate inputs identified are false positives, we employ the PI-detector~\cite{tan2025computingfloatingpointerrorsinjecting}, an efficient perturbation injection-based tool for floating-point error estimation, to identify and isolate true positives as the final results.

\begin{figure*}[htbp]
\centerline{\includegraphics[width=0.98\textwidth]{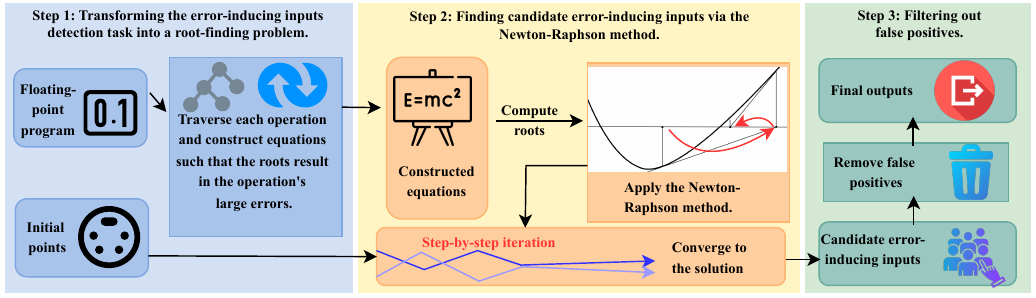}}
\caption{An overview of MGDE.}
\vspace{-0.3cm}
\label{framework}
\end{figure*}

In this section, we begin by presenting the theoretical foundation and the key design motivations underlying our proposed method. We then provide a detailed description of MGDE, which consists of three primary steps. We utilize the example in subsection~\ref{condition_number} to illustrate our method and show how false positives are introduced during the detection process. We further show that our method is capable of efficiently uncovering error-inducing inputs that FPCC fails to detect on \textit{gsl\_sf\_airy\_Ai\_deriv\_e} (cf. Section~\ref{motivate}). Figure~\ref{framework} provides an overview of our proposed method.

\subsection{Foundations of MGDE}
Before introducing the details of our method, this subsection presents its theoretical foundation---namely, that the root cause of floating-point errors lies in atomic operations with large condition numbers. We then explain how this insight motivates the use of the Newton-Raphson method for detecting error-inducing inputs.

\noindent \textbf{Theoretical foundation: Floating-point errors stem from atomic operations characterized by large condition numbers.} Atomic operations, including basic arithmetic (addition, subtraction, multiplication, division) as well as functions such as \(\log\) and \(\sin\), can be a source of significant floating-point errors. In subsection~\ref{condition_number}, we present an example where the expression \(\sin(x) - 0.4\) exhibits a large error when \(x\) is 0.411516846067, due to the large condition number of the subtraction operation. 

\noindent \textbf{Design motivation: How to develop a search algorithm that yields accurate directional guidance?} We anticipate the utilization of algorithms analogous to gradient descent in machine learning, enabling iterative convergence toward the optimum point, rather than aimless trial-and-error approaches. Hence, we must first define an objective function whose minimization directly produces the target error-inducing inputs. We draw design inspiration from an existing methodology known as ATOMU. ATOMU~\cite{zou2019detecting} leverages condition numbers of atomic operations to detect error-inducing inputs, thereby avoiding the need for time-consuming high-precision computations. Specifically, ATOMU employs an evolutionary algorithm to identify inputs that cause any operation within the function to exhibit a large condition number. However, this search process lacks clear guidance, making it akin to finding a needle in a haystack, and thus limits its effectiveness in detecting errors. We observe that when the condition number of a particular atomic operation is sufficiently large, the operation's output tends to converge to either a fixed value or infinity. Therefore, if a certain input causes the result of the operation to approach the particular value, the condition number of the operation becomes large, which in turn can lead to a significant error in the final result. For the expression \(\sin(x) - 0.4\), if the result approaches zero, the subtraction operation illustrates a large condition number, which can lead to a significant error in the output. Thus, we can directly set the equation as \(\sin(x) - 0.4=0\) and use the Newton-Raphson method to find the optimal value of \(x\) (i.e., the root of the equation). However, we cannot guarantee convergence to the desired solution for every initial point. Therefore, we need to initialize the method from multiple starting points.

\noindent \textbf{Why is gradient descent not selected as our strategy?} Gradient descent is a first-order algorithm~\cite{calafiore2014optimization}, while the convergence of the Newton-Raphson method is quadratic.

\subsection{Workflow of MGDE}

In this subsection, we introduce the details of our mathematically guided error-inducing detection method. We use the example in subsection~\ref{condition_number} to demonstrate both our detection methodology and the rationale necessitating the implementation of the third step. Moreover, in our presentation of Step 2, we show that our method can readily identify error-inducing inputs in \textit{gsl\_sf\_airy\_Ai\_deriv\_e}---error-inducing inputs that are missed by FPCC, the current state-of-the-art tool.

\noindent \textbf{Step 1 (addressing limitation 1 ): Transforming the error-inducing inputs detection task into a root-finding problem.} As stated in the theoretical foundation, the error in the output of a floating-point program originates from atomic operations with large condition numbers. Building on this principle, ATOMU~\cite{zou2019detecting} employs an evolutionary algorithm to identify inputs that lead to exceptionally large condition numbers for certain operations. To endow our method with long-range convergence, we reformulate the error-inducing detection problem as a root-finding task and employ the famous Newton-Raphson method. When the condition number of a certain atomic operation is very large, the result of this operand has two possible outcomes: one tends toward a specific value, and the other tends toward infinity (cf. Table 2 in ATOMU~\cite{zou2019detecting}). For the first case, such as the subtraction operation, our method for constructing the equation is to set the result of this operation to that value---for subtraction, this value is 0. For the second case, such as the \(\sinh(x)\) function, we set the reciprocal of the result of this operation to 0 and construct the equation.

In the first step, we select a set of initial points and apply the aforementioned transformation approach to traverse all operations and construct corresponding equations for each of them. Since we do not rely on high-precision programs to build the indicator used for detection, this addresses the first limitation.

\noindent \textbf{Step 2 (addressing limitation 2) : Finding candidate error-inducing inputs via the Newton-Raphson method.} The second step involves computing the roots of all equations constructed in the first step using the Newton-Raphson method, which is an efficient algorithm exhibiting quadratic convergence (cf. subsection~\ref{Newton-Raphson_method}). The iterative procedure of the Newton-Raphson method is conceptually simple and easy to implement: \(x_n = x_{n-1} - f(x_{n-1})/f'(x_{n-1})\), where \(x_{n-1}\) denotes the current iteration and \(x_n\) represents the next step. We do not need to specify a step size as required in gradient descent methods; instead, we only need to define a stopping criterion, such as terminating the process when the derivative falls below a certain threshold. When the function is treated as a black box (in our implementation), obtaining its analytical derivative becomes impractical. Thus, we compute the numerical derivative using the difference approximation: \((f(x + h) - f(x - h)) / (2 * h)\). The potential errors introduced by this numerical approximation are negligible, as an exact root is not required---an adequately close approximation value is sufficient to identify the desired error-inducing inputs.

\begin{figure}[htbp]
\vspace{-0.3cm}
    \centering
    \includegraphics[width=0.45\textwidth]{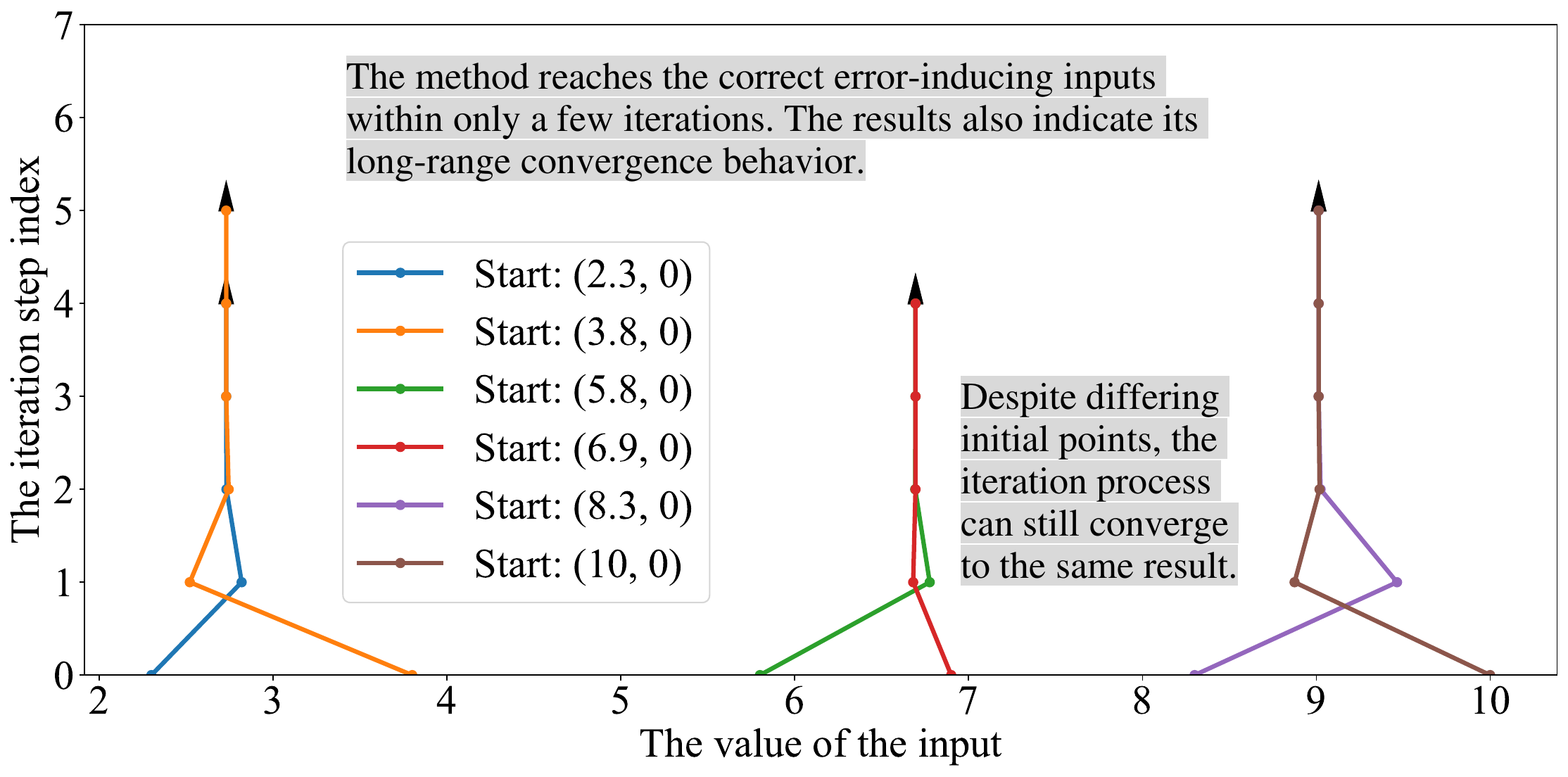}
    \vspace{-0.4cm}
    \caption{Convergence path of our method for the example. Arrows indicate convergence to the correct inputs.}
    \vspace{-0.2cm}
    \label{newton_paths}
\end{figure}

To demonstrate the search performance of our method, we employ \(\sin(x) - 0.4\) as a representative example. We focus on the subtraction in this expression, aiming to find the value of \(x\) that makes the result equal to zero. The \(\sin\) operation will be examined in the subsequent step. Figure~\ref{newton_paths} illustrates the efficiency and long-range convergence of our method. Despite starting from a distant initial point, MGDE still successfully converges to the correct error-inducing inputs.

\begin{figure*}[htbp]
\vspace{-0.3cm}
    \centering
    \includegraphics[width=0.95\textwidth]{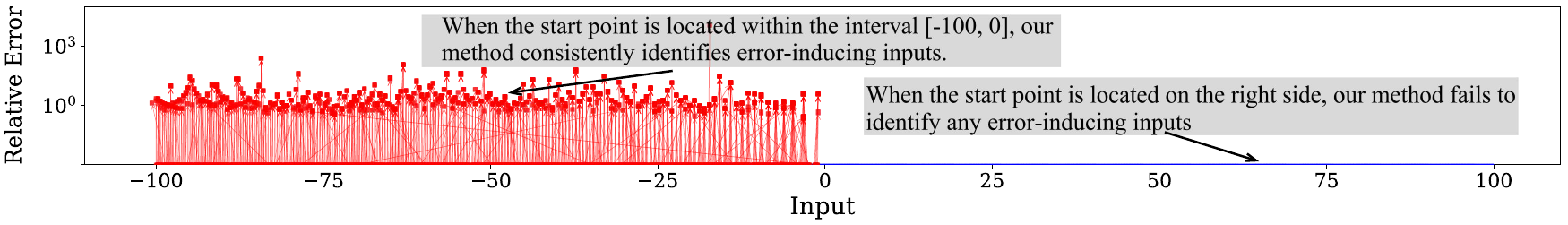}
    \vspace{-0.4cm}
    \caption{Detection performance of MGDE on \textit{gsl\_sf\_airy\_Ai\_deriv\_e}. The red dots on the coordinate axis represent that error-inducing inputs can be found using these points as start points, with the red arrows indicating the locations of the identified inputs and their corresponding relative errors. The blue dots denote cases where no such inputs could be detected.}
    \vspace{-0.2cm}
    \label{func_4}
\end{figure*}

To demonstrate the effectiveness of our method on complex functions, we apply it to \textit{gsl\_sf\_airy\_Ai\_deriv\_e}, where FPCC fails to identify any correct error-inducing inputs (cf. Section~\ref{motivate}). We construct 2,000 points at intervals of 0.1 within the range \([-100, 100]\) and apply our method using each as an initial start point. Figure~\ref{func_4} illustrates that 953 of these points successfully led to the identification of error-inducing inputs. This outcome underscores the robust long-range convergence capability of our method, enabling it to locate error-inducing inputs with minimal initial sampling and iterative attempts. As MGDE exhibits the long-range convergence properties, we are able to address the limitation 2.

\noindent \textbf{Step 3: Filtering out false positives.} In the final step, we need to eliminate the false positives generated in the second step. We select all inputs that lead to significant errors in a certain operation within the program; however, these inputs do not necessarily lead to large errors in the final program output, as subsequent operations may mitigate the initial error. In our example expression \(\sin(x) - 0.4\), the \(\sin(x)\) operation can incur errors as \(x\) approaches integer multiples of \(\pi\). For example, when \(x\) is equal to 3.14159265358979, the relative error of \(\sin(x)\) is \(2.28 \times 10^{-3}\), while the error of the final result is only \(1.03 \times 10^{-16}\). 

We use PI-detector~\cite{tan2025computingfloatingpointerrorsinjecting}, which is an efficient method to calculate the errors of floating-point programs, to eliminate false positives. PI-detector perturbs the program by injecting perturbations into operations with large condition numbers, then computes the final program error as the difference between the outputs of the original and perturbed versions. This approach effectively suppresses false positives, as errors introduced in individual operations---if later canceled out by subsequent computations---do not propagate to the final result of the perturbed program.
\section{Evaluation}
\label{eva}

In this section, we carry out empirical experiments to examine the effectiveness and efficiency of our method. We begin by presenting the implementation of our proposed method, followed by a detailed description of the experimental setup, including the datasets, evaluation metrics, and software/hardware configurations. Finally, we present and discuss the evaluation results. Since FPCC~\cite{yi2024fpcc} represents the state-of-the-art method for error-inducing input detection, we adopt it as our primary baseline for comparison. We also provide a case study (one GSL function) demonstrating that a single bug can be triggered by multiple inputs.

\subsection{Implementation}

We treat the target floating-point program as a black-box function in our analysis. First, the program is compiled into LLVM Intermediate Representation (IR)~\cite{lattner2004llvm}, a low-level, architecture-independent code representation. Next, we employ a custom LLVM pass to instrument the IR, enabling runtime data collection during program execution. Our instrumentation process systematically traverses all instructions in the IR. When encountering a floating-point operation, the pass inserts a call to an external handler. This handler extracts and records detailed information about the operation, including its operands and execution context. The collected data subsequently facilitates the application of the Newton-Raphson method for error-inducing input detection.

We implement the Newton-Raphson method in C++. We set a maximum of 20 iterations and define three convergence criteria as follows: \(|f(x)|<1\times10^{-15}\), \(|f'(x)|<1\times10^{-10}\), and \(|f(x)/f'(x)|<1\times10^{-10}\).

We partition the entire IEEE-754 double-precision input domain into 32 consecutive intervals. Since our method supports long-range convergence, we set sparser interval partitioning compared to FPCC. These intervals span from the smallest representable double value ($-1.8 \times 10^{308}$) to the largest ($1.8 \times 10^{308}$). The intervals are unevenly distributed: finer-grained subdivisions are applied around zero and coarser-grained bins are used in the extreme ranges. Specifically, the values of the 17 endpoints corresponding to the 16 intervals in the positive number domain are as follows: 0, \(1\times 10^{-100}\), 0.25, 0.5, 1.0, 2.0, 4.0, 8.0, 16.0, 32.0, \(1\times 10^{5}\), \(1\times 10^{8}\), \(1\times 10^{11}\), \(1\times 10^{14}\), \(1\times 10^{17}\), \(1\times 10^{20}\), and \(1.8\times 10^{308}\). The intervals on the negative axis are symmetrically related to them.

\subsection{Experiment setup}
\label{Experiment_setup}

\subsubsection{Dataset}

We use the dataset from our baseline method, FPCC, as our validation set. This dataset consists of 88 single-input functions from the GSL, an open-source numerical computing library that provides a wide range of mathematical algorithms. The GSL functions serve as a widely adopted dataset for error-inducing input detection, including LSGA~\cite{zou2015genetic}, DEMC~\cite{yi2019efficient}, and ATOMU~\cite{zou2019detecting}. 

In line with previous studies~\cite{zou2019detecting, yi2024fpcc}, we employ \textit{mpmath}~\cite{mpmath}, a library with arbitrary precision, to implement a high-precision version that serves as an oracle for validating the correctness of the identified error-inducing inputs, although our method itself does not depend on high-precision computation.

\subsubsection{Evaluation metric}

\textbf{Number of triggered bugs:} Consistent with previous studies~\cite{zou2019detecting, yi2024fpcc}, we consider relative errors with a magnitude exceeding 0.001 as significant errors, and maintain that such cases should be identified as error-inducing inputs. The primary objective of identifying error-inducing inputs is to assist developers in locating the triggered bugs---specifically, the problematic lines of code---which can subsequently be addressed through debugging efforts. Crucially, when multiple error-inducing inputs trigger identical bugs, their quantity becomes meaningless. Therefore, in our study, we mainly employ the number of distinct bugs triggered by the detected error-inducing inputs as our evaluation metric, rather than simply quantifying the error-inducing inputs themselves.

\textbf{Exclusion due to GSL's estimation error:} It is worth noting that GSL functions not only return the computed results but also provide an estimate of the associated relative error. Therefore, we argue that if the error-inducing inputs identified by a search tool correspond to large relative errors already reported by the GSL functions themselves, the findings are of limited significance. This is because the GSL functions are inherently aware of the substantial errors in these regions. Therefore, in our experiments, we eliminate error-inducing inputs for which GSL itself could identify the errors. We conduct experiments to assist in determining the threshold for our elimination. Specifically, since only a small fraction of inputs would cause large errors in floating-point programs~\cite{bao2013fly}, we randomly generate 10,000 double-precision floating-point numbers and execute each of the 88 GSL functions on them. Based on our results, 87\% of the relative errors are less than or equal to \(1 \times 10^{-12}\). Therefore, we set a relaxed threshold of \(1\times10^{-6}\): if the estimated error reported by GSL is greater than or equal to \(1\times10^{-6}\), we consider the bug to be detectable by GSL and can be discarded.

\subsubsection{Software and hardware environment}
Our primary experimental environment is a Docker container running Ubuntu 24.04 on a desktop equipped with an AMD Ryzen 5 9600X @ 3.90 GHz CPU and 32GB RAM.

\subsection{Evaluation results}

This subsection details the experimental results of our method, MGDE, structured around two research questions (RQs). For each RQ, we begin by outlining its motivation and evaluation methodology, then provide an in-depth discussion of the results.

\subsection*{RQ1: How effective is MGDE in detecting bugs of floating-pint programs? (Addressing limitation 2)} 
\label{rques1}
\noindent\textit{\textbf{Motivation.}} Errors in floating-point programs can lead to severe issues, with only a small subset of inputs capable of triggering significant errors. Consequently, identifying such inputs is of critical importance. Since multiple inputs may induce the same bug, merely comparing the quantity of error-inducing inputs discovered is not meaningful. Therefore, to validate the effectiveness of our method, we evaluate the number of distinct bugs identified in the dataset and compare the results with FPCC.

\noindent\textit{\textbf{Approach.}} We apply our method to the FPCC dataset, which consists of 88 single-input GSL functions. To demonstrate the long-range convergence behavior of MGDE, we generate only one random double-precision floating-point number as the initial point within each interval of our partitioning scheme. Since our initial results may contain false positives, we subsequently use PI-detector~\cite{tan2025computingfloatingpointerrorsinjecting}, based on perturbation injections, to efficiently compute the error and filter these false positives. We discard cases where the error can already be detected by GSL (cf. subsection~\ref{Experiment_setup}). For fairness, we employ the identical filtering process on the results from FPCC. Finally, we compute the relative error using high-precision programs and validate the detected error-inducing inputs with a threshold of 0.001. 

Our strategy for selecting bugs triggered by error-inducing inputs is grounded in the design principles of ATOMU, FPCC, and PI-detector. Specifically, large output errors in floating-point programs are caused by operations with large condition numbers. However, not all such operations necessarily lead to substantial errors in the final result. Followed by PI-detector, we first identify operations associated with error-inducing inputs that have condition numbers greater than \(1 \times 10^{5}\). We then individually perturb these operations and compute the relative error of the final output. If the error exceeds \(1 \times 10^{-10}\), we consider the operation a triggered bug. To ensure that our results are not affected by randomness, we report the average outcome over 100 independent runs.

\noindent\textit{\textbf{Results. }}\textbf{MGDE detects 85.42\% more bugs than FPCC and covers 51.72\% more functions.} Table~\ref{detection_result} illustrates that FPCC detects 380 error-inducing inputs, whereas our method identifies only 219. This discrepancy arises because MGDE supports long-range convergence, necessitating fewer initial sampling points and consequently yielding fewer detected inputs. However, since multiple inputs may trigger the same bug, we take the number of triggered bugs as the evaluation metric. Our method successfully uncovers 89 distinct bugs spanning 44 functions. In contrast, FPCC detects only 48 bugs across 29 functions. This limitation can be attributed to FPCC's capability of only achieving short-range convergence, thereby restricting its focus to easily detectable bugs. Our method demonstrates precise error-inducing input detection, identifying only 2 and 1 problematic inputs for functions \(gsl\_sf\_lambert\_W0\_e\) and \(gsl\_sf\_lambert\_Wm1\_e\) respectively. By comparison, FPCC returns 62 and 31 inputs for the same functions. Consequently, both methods ultimately detect just one actual bug in each function. Notably, all the bugs triggered by the error-inducing inputs found by FPCC are also identified by our method.

\begin{table*}[]
\caption{Detection results of our method and FPCC. (Speedup is defined as the execution time of FPCC divided by that of our method. Neither method employs parallel optimization during the experiments.)}
\small
\label{detection_result}
\begin{tabular}{lccccrrrrc}
\toprule
\multicolumn{1}{c}{\multirow{2}{*}{\textbf{GSL Functions}}} & \multicolumn{2}{c}{\textbf{\# Error-triggering Inputs}}                  & \multicolumn{2}{c}{\textbf{\# Triggered Bugs}}                           & \multicolumn{3}{c}{\textbf{Time (s)}}                                                                         & \multicolumn{2}{c}{\textbf{\# Triggered Bugs Per Second}}                \\
\cmidrule(lr){2-3} \cmidrule(lr){4-5} \cmidrule(lr){6-8} \cmidrule(lr){9-10}
\multicolumn{1}{c}{}                                        & \multicolumn{1}{c}{\textbf{MGDE}} & \multicolumn{1}{c}{\textbf{FPCC}} & \multicolumn{1}{c}{\textbf{MGDE}} & \multicolumn{1}{c}{\textbf{FPCC}} & \multicolumn{1}{c}{\textbf{MGDE}} & \multicolumn{1}{c}{\textbf{FPCC}} & \multicolumn{1}{c}{\textbf{Speedup}} & \multicolumn{1}{c}{\textbf{MGDE}} & \multicolumn{1}{c}{\textbf{FPCC}} \\
\midrule
gsl\_sf\_airy\_Ai\_e                                        & 3                                 & 0                                 & 2                                 & 0                                 & 0.0262                            & 0.1535                            & 5.8655                                & 75                                & 0                                 \\
gsl\_sf\_airy\_Bi\_e                                        & 3                                 & 0                                 & 2                                 & 0                                 & 0.0262                            & 0.1429                            & 5.4648                                & 91                                & 0                                 \\
gsl\_sf\_airy\_Ai\_scaled\_e                                & 3                                 & 0                                 & 2                                 & 0                                 & 0.0273                            & 0.1595                            & 5.8440                                & 72                                & 0                                 \\
gsl\_sf\_airy\_Bi\_scaled\_e                                & 3                                 & 0                                 & 2                                 & 0                                 & 0.0284                            & 0.1624                            & 5.7150                                & 75                                & 0                                 \\
gsl\_sf\_airy\_Ai\_deriv\_e                                 & 5                                 & 0                                 & 1                                 & 0                                 & 0.0079                            & 0.0488                            & 6.1919                                & 127                               & 0                                 \\
gsl\_sf\_airy\_Bi\_deriv\_e                                 & 3                                 & 0                                 & 2                                 & 0                                 & 0.0088                            & 0.0499                            & 5.6711                                & 214                               & 0                                 \\
gsl\_sf\_airy\_Ai\_deriv\_scaled\_e                         & 5                                 & 0                                 & 1                                 & 0                                 & 0.0093                            & 0.0582                            & 6.2463                                & 107                               & 0                                 \\
gsl\_sf\_airy\_Bi\_deriv\_scaled\_e                         & 3                                 & 0                                 & 2                                 & 0                                 & 0.0103                            & 0.0655                            & 6.3560                                & 181                               & 0                                 \\
gsl\_sf\_bessel\_J0\_e                                      & 11                                & 6                                 & 4                                 & 2                                 & 0.0279                            & 0.2055                            & 7.3636                                & 150                               & 10                                \\
gsl\_sf\_bessel\_J1\_e                                      & 11                                & 8                                 & 4                                 & 2                                 & 0.0276                            & 0.2035                            & 7.3714                                & 148                               & 10                                \\
gsl\_sf\_bessel\_Y0\_e                                      & 5                                 & 4                                 & 3                                 & 2                                 & 0.0143                            & 0.0900                            & 6.3148                                & 212                               & 22                                \\
gsl\_sf\_bessel\_Y1\_e                                      & 5                                 & 4                                 & 3                                 & 2                                 & 0.0137                            & 0.0888                            & 6.4820                                & 229                               & 23                                \\
gsl\_sf\_bessel\_j0\_e                                      & 17                                & 0                                 & 1                                 & 0                                 & 0.0004                            & 0.0074                            & 20.1756                               & 2,727                             & 0                                 \\
gsl\_sf\_bessel\_j1\_e                                      & 15                                & 4                                 & 2                                 & 1                                 & 0.0016                            & 0.0109                            & 6.9920                                & 1,283                             & 92                                \\
gsl\_sf\_bessel\_j2\_e                                      & 14                                & 4                                 & 2                                 & 1                                 & 0.0014                            & 0.0161                            & 11.5633                               & 1,441                             & 62                                \\
gsl\_sf\_bessel\_y0\_e                                      & 5                                 & 20                                & 3                                 & 3                                 & 0.0023                            & 0.0577                            & 24.7933                               & 1,277                             & 52                                \\
gsl\_sf\_bessel\_y1\_e                                      & 7                                 & 24                                & 5                                 & 4                                 & 0.0061                            & 0.1233                            & 20.1329                               & 771                               & 32                                \\
gsl\_sf\_bessel\_y2\_e                                      & 7                                 & 20                                & 5                                 & 4                                 & 0.0054                            & 0.1279                            & 23.4915                               & 903                               & 31                                \\
gsl\_sf\_clausen\_e                                         & 4                                 & 16                                & 3                                 & 2                                 & 0.0253                            & 0.0756                            & 2.9926                                & 112                               & 26                                \\
gsl\_sf\_dilog\_e                                           & 3                                 & 1                                 & 1                                 & 1                                 & 0.0073                            & 0.0690                            & 9.4837                                & 138                               & 15                                \\
gsl\_sf\_expint\_E1\_e                                      & 1                                 & 1                                 & 1                                 & 1                                 & 0.0142                            & 0.0660                            & 4.6454                                & 70                                & 15                                \\
gsl\_sf\_expint\_E2\_e                                      & 4                                 & 1                                 & 1                                 & 1                                 & 0.0146                            & 0.0674                            & 4.6168                                & 68                                & 15                                \\
gsl\_sf\_expint\_E1\_scaled\_e                              & 1                                 & 1                                 & 1                                 & 1                                 & 0.0273                            & 0.1618                            & 5.9163                                & 37                                & 6                                 \\
gsl\_sf\_expint\_E2\_scaled\_e                              & 3                                 & 0                                 & 1                                 & 0                                 & 0.0228                            & 0.1604                            & 7.0427                                & 44                                & 0                                 \\
gsl\_sf\_expint\_Ei\_e                                      & 1                                 & 1                                 & 1                                 & 1                                 & 0.0156                            & 0.0624                            & 3.9999                                & 64                                & 16                                \\
gsl\_sf\_expint\_Ei\_scaled\_e                              & 1                                 & 1                                 & 1                                 & 1                                 & 0.0258                            & 0.1647                            & 6.3819                                & 39                                & 6                                 \\
gsl\_sf\_Chi\_e                                             & 4                                 & 2                                 & 1                                 & 1                                 & 0.0383                            & 0.1454                            & 3.7939                                & 26                                & 7                                 \\
gsl\_sf\_Ci\_e                                              & 9                                 & 25                                & 6                                 & 4                                 & 0.0218                            & 0.1921                            & 8.8241                                & 262                               & 21                                \\
gsl\_sf\_lngamma\_e                                         & 4                                 & 2                                 & 2                                 & 1                                 & 0.0066                            & 0.0414                            & 6.2366                                & 337                               & 24                                \\
gsl\_sf\_lambert\_W0\_e                                     & 2                                 & 62                                & 1                                 & 1                                 & 0.0012                            & 0.0259                            & 21.0769                               & 813                               & 39                                \\
gsl\_sf\_lambert\_Wm1\_e                                    & 1                                 & 31                                & 1                                 & 1                                 & 0.0011                            & 0.0260                            & 23.4405                               & 901                               & 38                                \\
gsl\_sf\_legendre\_P2\_e                                    & 4                                 & 2                                 & 1                                 & 1                                 & 0.0007                            & 0.0065                            & 9.8392                                & 1,516                             & 154                               \\
gsl\_sf\_legendre\_P3\_e                                    & 4                                 & 2                                 & 1                                 & 1                                 & 0.0008                            & 0.0090                            & 10.9869                               & 1,226                             & 112                               \\
gsl\_sf\_legendre\_Q1\_e                                    & 2                                 & 2                                 & 1                                 & 1                                 & 0.0004                            & 0.0139                            & 35.2838                               & 2,521                             & 72                                \\
gsl\_sf\_log\_e                                             & 1                                 & 0                                 & 1                                 & 0                                 & 0.0001                            & 0.0051                            & 47.9050                               & 9,475                             & 0                                 \\
gsl\_sf\_log\_abs\_e                                        & 3                                 & 0                                 & 1                                 & 0                                 & 0.0004                            & 0.0052                            & 11.9230                               & 2,301                             & 0                                 \\
gsl\_sf\_psi\_e                                             & 7                                 & 0                                 & 4                                 & 0                                 & 0.0230                            & 0.1374                            & 5.9852                                & 191                               & 0                                 \\
gsl\_sf\_sin\_e                                             & 5                                 & 0                                 & 2                                 & 0                                 & 0.0065                            & 0.0927                            & 14.2261                               & 307                               & 0                                 \\
gsl\_sf\_cos\_e                                             & 6                                 & 0                                 & 2                                 & 0                                 & 0.0053                            & 0.0866                            & 16.3631                               & 378                               & 0                                 \\
gsl\_sf\_sinc\_e                                            & 14                                & 126                               & 4                                 & 4                                 & 0.0050                            & 0.0986                            & 19.6484                               & 797                               & 41                                \\
gsl\_sf\_lnsinh\_e                                          & 2                                 & 1                                 & 1                                 & 1                                 & 0.0002                            & 0.0066                            & 33.6009                               & 5,104                             & 152                               \\
gsl\_sf\_zeta\_e                                            & 3                                 & 4                                 & 1                                 & 1                                 & 0.0246                            & 0.0899                            & 3.6533                                & 40                                & 11                                \\
gsl\_sf\_zetam1\_e                                          & 2                                 & 1                                 & 2                                 & 1                                 & 0.0245                            & 0.0791                            & 3.2310                                & 73                                & 13                                \\
gsl\_sf\_eta\_e                                             & 3                                 & 4                                 & 1                                 & 1                                 & 0.0280                            & 0.0984                            & 3.5163                                & 36                                & 10                                \\  \cmidrule(lr){1-5}
Total                                                       & 219                               & 380                               & 89                                & 48                                &           &                    &               & \multicolumn{1}{l}{}              & \multicolumn{1}{l}{}                  \\  \cmidrule(lr){6-10}
Average   & & & &  & 0.0133      & 0.0854       & 6.4096   & 839.9773 & 25.6136 \\
\bottomrule
\end{tabular}
\end{table*}

\noindent\textbf{The true positive rate of the error-inducing inputs detected by our method is 97.80\%.} The error-inducing inputs reported in Table~\ref{detection_result} are validated using the oracles implemented with \textit{mpmath}. The original results produced by our method exhibit a true positive rate of 97.80\%. Moreover, the majority of false positives have relative error values close to 0.001, indicating that the error-inducing inputs identified by our method are highly reliable.

\findingboxx{Our method is effective in identifying bugs that lead to substantial errors in programs, demonstrating the practical value.
}


\subsection*{RQ2: How efficient is MGDE in detecting bugs of floating-pint programs? (Addressing limitation 1)}

\noindent\textit{\textbf{Motivation.}} In RQ1, we demonstrate that our method can effectively detect a substantial number of bugs. In practical applications, the runtime of the tool is also a critical metric, as excessively slow execution may discourage developers from adopting such tools. Theoretically, our method achieves shorter runtime due to its support for long-range convergence, which reduces the number of required probes. To validate the efficiency of MGDE, we record its runtime and conduct a comparative analysis with FPCC.

\noindent\textit{\textbf{Approach.}} The experimental procedure for RQ2 is identical to that of RQ1. To mitigate the impact of randomness, all results are obtained by averaging over 100 executions. We apply the Mann-Whitney-Wilcoxon test~\cite{wilcoxon1992individual} to determine whether the differences between our method and FPCC are statistically significant.

\noindent\textit{\textbf{Results. }}\textbf{The total running time of FPCC is 6.4096 times that of our proposed method.} Our method achieves an average runtime of 0.0133 seconds across 44 functions, while FPCC requires 0.0854 seconds. To validate the statistical significance of the runtime difference between the two methods, we employ the Mann-Whitney-Wilcoxon test. The experimental results yield a p-value of \(1.2035 \times 10^{-9}\), demonstrating a statistically significant difference between the two methods. These findings indicate that our method substantially outperforms FPCC in terms of computational efficiency. Consequently, our method exhibits greater practical applicability, particularly for complex large-scale programs where single execution times are inherently prolonged.

\noindent\textbf{With respect to the average number of bugs identified per function per second, our method achieves a 32.7941× speedup over FPCC.} On average, our method is able to detect 839.9773 bugs per function per second, whereas FPCC detects only 25.6136. These results demonstrate that MGDE is not only effective in identifying bugs, but also highly efficient. In some cases, it can even uncover bugs missed by FPCC in less than 0.0001 seconds.

\findingboxx{Our method shows remarkable efficiency, exhibiting a 32.7941× speedup (bugs detected per second) compared to the baseline algorithm. This significant advantage enables broader application prospects, particularly in more complex and long-running programs.}

\subsection{Evaluating MGDE with other general quality metrics}
In RQ1 and RQ2, we empirically validate both the effectiveness and efficiency of our proposed method. It should be noted, however, that MGDE is inherently non-deterministic due to its reliance on randomly generated initial sampling points. Furthermore, real-world detection applications often involve functions with multiple parameters. To address these practical considerations, this subsection evaluates the stability and scalability of our method under these more complex conditions.

To examine whether MGDE can stably detect a substantial number of bugs, we conduct multiple runs of the experiments and analyze the stability of the results. We independently execute our method 100 times, recording the number of error-inducing inputs and triggered bugs in each run. The stability of the 100 experimental results is then analyzed. Table~\ref{100_runs} shows that even in the worst-case scenario, our method is capable of identifying 202 error-inducing inputs and the corresponding 83 bugs. It is worth noting that FPCC detects only 48 bugs, indicating that our method still significantly outperforms FPCC even in the worst condition. Furthermore, across 100 independent experimental runs, our method consistently detects the same set of 71 common bugs, demonstrating its ability to reliably identify these bugs regardless of the initial sampling distribution. In contrast, FPCC only identifies 48 bugs in total.

To assess our method's scalability, we evaluate its performance on multi-input functions using 18 dual-input GSL functions as benchmark cases, with FPCC serving as the baseline comparison. We intentionally exclude FPCC's native multi-input function support from this evaluation because while such functions exhibit high input dimensionality, they are typically simple (e.g., \(recursive\_summation\)). In local operations such as addition and subtraction, numerous distinct errors can be easily generated. Consequently, identifying these error-inducing inputs provides limited practical value. In contrast, similar to our single-input dataset, these dual-input GSL functions are professionally implemented mathematical routines with high reliability. Consequently, the detection of errors in these functions serves as a strong indicator of our method's effectiveness. During the experimental phase, we adapt the Newton-Raphson method to its multivariate form to accommodate our method. In contrast, FPCC handles multi-input cases differently, as it lacks a unified interface for multivariate functions. To implement FPCC for comparison, we extract the source code of these 18 dual-input GSL functions and integrate them directly into FPCC's framework. We maintain all original FPCC parameters without modification. However, we acknowledge that the absence of a native calling interface and unmodified parameters may have prevented FPCC from achieving its full potential in this evaluation. Our method detects one, two, five, and one bugs in \(gsl\_sf\_bessel\_Ynu\_e\), \(gsl\_sf\_bessel\_Jn\_e\), \(gsl\_sf\_bessel\_Yn\_e\), and \(gsl\_sf\_beta\_e\), respectively. Conversely, FPCC fails to detect any bugs across the 18 functions. Moreover, our method achieves an average execution time of 0.6443 seconds across all 18 GSL functions. In comparison, FPCC is constrained to a 100-second time budget per function. Despite this substantial computational allocation, FPCC fails to detect any bugs. These results demonstrate the superior scalability of our method in high-dimensional input spaces. While heuristic search algorithms suffer from exponential time complexity growth with increasing dimensions (due to required subspace partitioning), our method maintains efficiency through its long-range convergence property, enabling effective detection of error-inducing inputs and associated bugs. These results also demonstrate the potential for our method to serve as an effective supplementary tool in future testing pipelines, enhancing error detection capabilities in practical applications.

\findingboxx{Our method exhibits stable and scalable performance. This result indirectly reflects the advantage of MGDE in achieving long-range convergence, as it discovers bugs regardless of the initial inputs and input dimensions.}

\subsection{A case study}

\begin{table}[]
\caption{Statistical analysis of 100 execution results.}
\label{100_runs}
\begin{tabular}{lrrrr}
\toprule
& \multicolumn{1}{c}{\textbf{Max}} & \multicolumn{1}{c}{\textbf{Min}} & \multicolumn{1}{c}{\textbf{Average}} & \multicolumn{1}{c}{\textbf{Median}} \\
\midrule
\# Error-inducing inputs & 236 & 202 & 219.81 & 219.50\\
\# Triggered bugs &96 & 83 & 89.32 & 89.00 \\
\bottomrule
 \vspace{-0.7cm}
\end{tabular}
\end{table}

\begin{figure*}[htbp]
\vspace{-0.3cm}
    \centering
    \includegraphics[width=0.95\textwidth]{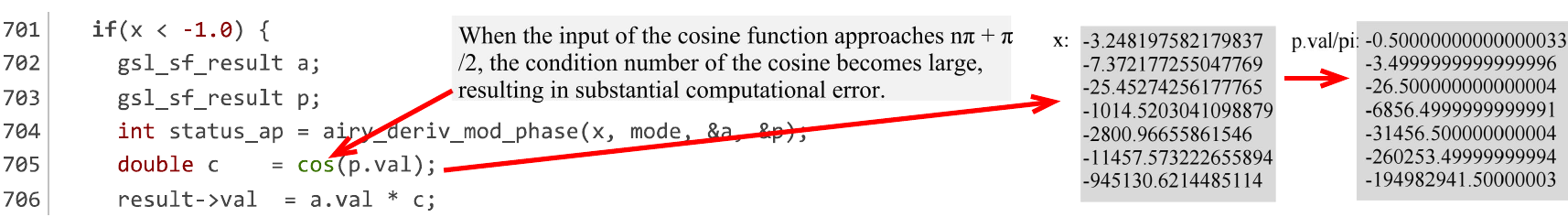}
    \caption{Code snippet of \textit{gsl\_sf\_airy\_Ai\_deriv\_e}. Seemingly unrelated variables \(x\) can simultaneously induce significant errors in cosine computations.}
    \vspace{-0.4cm}
    \label{code_snip}
\end{figure*}

This subsection presents an example to demonstrate that certain bugs can be triggered by multiple error-inducing inputs. The focus of our case study is the \textit{gsl\_sf\_airy\_Ai\_deriv\_e} function, which implements the Airy function derivative. One core part of this program is illustrated in Figure~\ref{code_snip}, where the \(\cos\) is located at line 705. According to the condition number formula, the condition number of the \(\cos(x)\) is given by: \(C_{\cos}(x)=|x\cdot\tan(x)|\). Therefore, if \(x\to n\pi+\frac\pi2,n\in\mathbb{Z}\), \(\cos(x)\) can result in large errors. Since the final result of the program is equal to the value of the \(\cos\) multiplied by a scalar, the error in the \(\cos\) is not canceled out and can lead to a large error in the final result. 

Figure~\ref{code_snip} demonstrates that numerous seemingly irregular inputs can result in large errors in the \(\cos\) operation, thereby leading to substantial inaccuracies in the program \textit{gsl\_sf\_airy\_Ai\_deriv\_e}. These error-inducing inputs trigger the same underlying bug; thus, identifying a single representative case suffices for analysis. This phenomenon highlights the superiority of our method, as it enables rapid convergence over long distances to identify an error-inducing input, thereby circumventing the need for exhaustive and inefficient needle-in-a-haystack searches. In practical applications, when users or developers need to determine whether a small input range may cause significant computational errors, our method can efficiently accomplish this task by randomly selecting only a few initial sampling points, rather than requiring dense probing across the search space. 

\section{Threats to Validity}
\label{threat}
This section discusses possible threats to the validity of our study.

\noindent \textbf{External validity.} A potential threat to external validity is that our method may only be effective within our specific dataset. However, the GSL functions dataset we employ has been widely utilized in prior research on error-inducing input detection, establishing its relevance and generalizability. Furthermore, the GSL is a highly popular and extensively adopted mathematical library. By successfully identifying error-inducing inputs in these functions, we demonstrate that our method is not limited to our experimental data but can also generalize well to other real-world scenarios. 

\noindent \textbf{Internal validity.} One potential threat is that the Newton-Raphson method does not guarantee convergence for all inputs. To mitigate this, we employ multiple initial points. Additionally, since only a small fraction of inputs induce significant errors---and many error-inducing inputs trigger the same underlying bug---our method remains effective in uncovering a comprehensive set of bugs.

\noindent \textbf{Construct validity.} A potential concern would arise if we solely compare methods based on the number of error-inducing inputs detected, as this metric alone is not meaningful. Since some bugs can be triggered by numerous inputs, discovering hundreds of such inputs may not provide more insight than identifying just one---in fact, more inputs may correspond to fewer unique bugs. To ensure a valid evaluation, our primary metric focuses on the number of distinct bugs triggered by the error-inducing inputs, rather than their raw quantity. 
\section{Related Work}
\label{related-work}
In this section, we first review prior works closely related to our research. We then discuss existing approaches that employ function minimization techniques for analyzing floating-point programs.

\noindent \textbf{\textit{Searching for error-inducing inputs.}} A growing body of research has focused on developing methods to detect inputs that cause significant floating-point errors. BGRT~\cite{chiang2014efficient} employs a heuristic-based search strategy to identify inputs that trigger substantial floating-point errors. Zou et al.~\cite{zou2015genetic} introduced the first metaheuristic search-based method for generating erroneous inputs. Yi et al.~\cite{yi2019efficient} presented an approach based on Differential Evolution algorithm~\cite{storn1997differential} and Monte Carlo Markov Chain algorithm (MCMC)~\cite{andrieu2003introduction}. RADE~\cite{wang2022detecting} integrates ranking analysis with search algorithms to guide the exploration process. The core component involves using three types of fitness functions to reduce the overall search time. Guo et al.~\cite{guo2020efficient} framed the problem as maximizing code coverage, which they addressed using symbolic execution. Zhang et al.~\cite{zhang2023eiffel} introduced EIFFEL, a novel error analysis framework designed for floating-point error prediction. The proposed methodology involves an initial sampling phase utilizing clustering algorithms to gather representative data points, followed by the application of polynomial fitting techniques to construct an approximate curve and predict errors. Zhang et al.~\cite{zhang2024hierarchical} designed a hierarchical search strategy, in which each layer corresponds to a specific precision level. Zou et al.~\cite{zou2019detecting} proposed ATOMU, a method for detecting significant errors by analyzing whether a given input could induce large atomic condition numbers in any arithmetic operation within the program. FPCC~\cite{yi2024fpcc} is the state-of-the-art algorithm that improves efficiency by using chain conditions as the objective function for input search, rather than relying on errors obtained from high-precision programs. Notably, it also addresses the false positive issue present in ATOMU.

Among the aforementioned tools, only ATOMU and FPCC do not rely on high-precision computations. Consequently, they achieve significantly higher efficiency by avoiding the computational overhead of high-precision programs. However, all existing tools---including ATOMU and FPCC---fail to employ efficient gradient-based optimization algorithms, resulting in suboptimal performance.

\noindent \textbf{\textit{Mathematical optimization for floating-point programs analysis.}} Fu et al.~\cite{fu2017achieving} solved the coverage-based testing problem by using unconstrained programming. They generated a representing function based on the floating-point program, where any minimum of this function corresponds to a test input that is certain to trigger a previously unexplored execution path in the program under test. Fu et al.~\cite{fu2016xsat} formalized the equivalence between floating-point satisfiability and unconstrained mathematical optimization. Fu et al.~\cite{fu2019effective} presented a systematic and theoretical reduction of floating-point analysis problems to mathematical optimization. Their methodology can be applied to key tasks such as boundary value analysis, path reachability and overflow detection.

Strictly speaking, the Newton-Raphson method is not technically a mathematical optimization algorithm. However, it offers superior quadratic convergence rates. By reformulating the error-inducing inputs detection task as a root-finding problem, we leverage the Newton-Raphson method's computational efficiency to identify these critical inputs. This novel adaptation endows our method with long-range convergence capabilities---a key advantage absent in all prior methods.
\section{Conclusion}
\label{conclusion}

Detecting error-inducing inputs has become an increasingly active research topic in software engineering and programming languages, as floating-point programs are foundational to modern society, and such inputs can help developers uncover bugs hidden within these foundations. Many existing approaches rely on high-precision computations to generate error indicators for their search procedures. However, such high-precision methods suffer from significant computational inefficiency, requiring substantial resources to execute. Moreover, all existing tools fail to formulate the search process as an algorithm that supports rapid iteration and long-range convergence, making the search akin to finding a needle in a haystack. This leads to highly inefficient search processes that frequently miss critical bugs. Furthermore, for multi-input scenarios, these methods exhibit exponential time complexity growth, making them impractical for real-world applications. To address this critical challenge, we reformulate error-inducing input detection as a root-finding problem and adopt the Newton-Raphson method. Experimental results demonstrate that our method decisively outperforms the current state-of-the-art, both in terms of effectiveness and efficiency. We believe our method opens a new era for the detection task and paves the way for future search algorithms grounded in solid mathematical theory.

\bibliographystyle{ACM-Reference-Format}
\bibliography{sample-base}

\end{document}